\documentclass[12pt]{article}
\usepackage{times}
\usepackage{geometry}
\usepackage[utf8]{inputenc}
\usepackage{enumitem,amssymb}

\usepackage{graphicx}	
\usepackage{ragged2e}
\usepackage{subfiles}
\usepackage[dvipsnames,svgnames,x11names]{xcolor}
\usepackage[colorlinks=true, linktocpage, linkcolor={blue!40!black}, citecolor={blue!40!black}, urlcolor={blue!50!black}]{hyperref}
\newlist{thematic}{itemize}{8}
\setlist[thematic]{label=$\square$}
\usepackage[para]{footmisc}
\usepackage{amssymb}
\usepackage{wrapfig}
\usepackage{pifont}

\newcommand{\emphasize}[1]{{\bf #1}}  
\usepackage[numbers]{natbib}
\newcommand{\arcsec}{\hbox{$^{\prime\prime}$}}       
\newcommand{\micron}{$\mu$m}                      

\usepackage{amsmath,amssymb,amstext}

\usepackage[innercaption]{sidecap}
\sidecaptionvpos{figure}{c}

\begin{document}
\raggedright
\huge
Astro2020 APC White Paper \linebreak

The Atacama Large Aperture Submillimeter Telescope (AtLAST) \linebreak
\normalsize

\noindent \textbf{Type of Activity:}
\linebreak
$\blacksquare$ Ground Based Project
\hspace*{10pt} $\square$ Space Based Project
\hspace*{20pt} $\square$ Infrastructure Activity
\linebreak
$\square$ Technology Development Activity
$\square$ State of the Profession Consideration
\hspace*{1pt} $\square$ Other
\hspace*{65pt} 

\textbf{Panel:}  Radio, Millimeter, and Submillimeter Observations from the Ground.\\

\vspace{12pt}

\noindent \textbf{Thematic Areas:}
\linebreak
$\blacksquare$ Planetary Systems \hspace*{10pt} $\blacksquare$ Star and Planet Formation \hspace*{20pt}\linebreak
$\blacksquare$ Formation and Evolution of Compact Objects \hspace*{31pt} $\blacksquare$ Cosmology and Fundamental Physics \linebreak
  $\blacksquare$  Stars and Stellar Evolution \hspace*{1pt} $\square$ Resolved Stellar Populations and their Environments \hspace*{40pt} \linebreak
  $\blacksquare$    Galaxy Evolution   \hspace*{45pt} $\square$             Multi-Messenger Astronomy and Astrophysics \hspace*{65pt} \linebreak
  
\textbf{Principal Author:}

Name:	Dr.\ Pamela D. Klaassen
 \linebreak						
Institution:  UK Astronomy Technology Centre
 \linebreak
Email: pamela.klaassen@stfc.ac.uk
 \linebreak
Phone:   + 44 0131 6688 218
 \linebreak
 
\textbf{Co-authors:}
\linebreak
Dr.\ Tony Mroczkowski (European Southern Observatory -- ESO),
Dr.\ Sean Bryan (ASU),
Dr.\ Christopher Groppi (Arizona State University -- ASU),
Dr.\ Kaustuv Basu (University of Bonn),
Dr.\ Claudia Cicone (University of Oslo),
Dr.\ Helmut Dannerbauer (Instituto de Astrof\'{i}sica de Canarias, Universidad de La Laguna),
Dr.\ Carlos De Breuck (ESO),
Dr.\ William J. Fischer (Space Telescope Science Institute),
Dr.\ James Geach (University of Hertfordshire),
Dr.\ Evanthia Hatziminaoglou (ESO-Garching), 
Prof.\ Wayne Holland (UK Astronomy Technology Centre), 
Prof.\ Ryohei Kawabe (National Astronomical Observatory of Japan),
Dr.\ Neelima Sehgal (Stony Brook University \& Flatiron Institute),
Dr.\ Thomas Stanke (ESO),
Dr.\ Eelco van Kampen (ESO).
\linebreak

\noindent \textbf{Endorsers listed after Abstract}

\justify

\textbf{Abstract:}

The sub-mm sky is a unique window for probing the architecture of the Universe and structures within it. From the discovery of dusty sub-mm galaxies, to the ringed nature of protostellar disks, our understanding of the formation, destruction, and evolution of objects in the Universe requires a comprehensive view of the sub-mm sky. The current generation single-dish sub-mm facilities have given a glimpse of the potential for discovery, while sub-mm interferometers have presented a high resolution view into the finer details of regions/objects. However, our understanding of large-scale structure and our full use of these interferometers is now hampered by the limited sensitivity of our sub-mm view of the universe at larger scales.

Thus, now is the time to start planning the next generation of sub-mm single dish facilities - to build on these revolutions in our understanding of the sub-mm sky. Here we present the case for the Atacama Large Aperture Submillimeter Telescope (AtLAST), a concept for a 50m class single dish telescope. We envision AtLAST as a facility operating as an international partnership with a suite of instruments to deliver the transformative science described in many Astro2020 science white papers. 

A 50m telescope with a high throughput and 1$^\circ$ field of view  with a full complement of advanced instrumentation, including highly multiplexed high-resolution spectrometers, continuum cameras and Integral Field Units, AtLAST will have mapping speeds thousands of times greater than any current or planned facility. It will reach confusion limits below $L_*$ in the distant universe and resolve low-mass protostellar cores at the distance of the Galactic Center, providing synergies with upcoming facilities across the spectrum (from SKA, SPHEREx and SPICA to the LSST and eROSITA). Located on the Atacama plateau,  to observe frequencies un-obtainable by other observatories, AtLAST  will enable  a fundamentally new understanding of the sub-mm universe at unprecedented depths.

\raggedright

\vspace{12pt}
\textbf{Endorsers:} (Listed Alphabetically)
\linebreak   

F. C Adams (University of Michigan), J. Afonso (Institute of Astrophysics and Space Sciences), C. Agliozzo (European Southern Observatory), S. Amodeo (Cornell), P. Andreani (ESO), M. Aravena (Universidad Diego Portales), T. Bakx (Nagoya University), A. Baryshev (Kapteyn Astronomical Institute), N. Battaglia (Cornell University), V. Belitsky (Chalmers University of Technology), M. Beltran (INAF-Osservatorio Astrofisico di Arcetri)
% ...


\justify
\pagebreak
\setcounter{page}{1}
\section{Background and Motivation}

Submillimeter observations reveal the properties of the many phases of baryonic matter. Single dish telescopes are able to recover the larger scales of many cosmic structures; from the Solar System to galaxy clusters and cosmic filaments. A 50-meter class submillimeter telescope with a large, degree-scale field of view would allow fast surveys with low confusion limits, complementing powerful interferometers such as ALMA\footnote{\url{almaobservatory.org}} that deliver high-resolution, detailed studies on individual objects.

A wide-field photometric and spectroscopic survey of the sub-mm sky with such a large telescope would provide the definitive probe of galaxy evolution from Cosmic Dawn to the present.  It would also enable  tracing and following the cycling of warm and cold baryons in and out of galaxies on scales of galaxy clusters (e.g. several arcminutes).  Through observations of the Sunyaev-Zeldovich effect, it could provide a comprehensive view on the warm/hot, ionized gas in large scale structures such as galaxies, groups, and clusters.  More locally, the high spatial resolution and mapping power of such a telescope would enable unique studies of planetary systems, the variable nature of star formation, and understandings of the properties of the cool molecular gas, dust, magnetic fields, and the hot ionized gas in our Galaxy.

{\bf The Atacama Large Aperture Submillimeter Telescope (AtLAST, \cite{Bertoldi2018})\footnote{\url{atlast-telescope.org}} project will deliver a much needed combination of spatial and spectral resolution, high mapping speeds and sensitivity to large scale structures that current facilities cannot achieve}. Currently, AtLAST is an international effort to study the scientific merit for and possible technical implementation of such a large-aperture mm/sub-mm telescope, a powerful next generation single-dish to complement ALMA to serve the future facility needs of the world-wide sub-mm community  (covering 35 to 950 GHz, or 8 to 0.3mm). Such a telescope has been identified by both the ESO and ALMA boards as a development priority (\cite{Testi15,Carpenter19}), but beyond the scope of the ALMA trilateral agreement. We envision AtLAST to be an international partnership operating a facility telescope, with an instrument suite providing high-resolution, multi-beam spectroscopy, ultra-wideband wide-field spectroscopy for lower spectral resolution mm/sub-mm tomography, and ultra-wide-field multi-chroic polarimetric imaging capabilities. We expect this (dome-less) facility could be built within 10 years, and have an operational lifetime of at least 30.

Although ALMA can follow-up many of the discoveries from current facilities in exquisite detail, there is a crucial gap in survey capabilities at mm/sub-mm/far-IR wavelengths.  ALMA cannot survey for the unknown interesting objects, or for the generalized statistical samples needed in the era of the SKA/ngVLA and big data. For this, a large, wide-bandwidth single dish telescope with a large instantaneous field of view (FoV) is required. Here we outline the key science cases that can only be accomplished with an AtLAST-style telescope (\S~\ref{sec:KeyScience}), along with describing the telescope itself (\S~\ref{sec:TechOverview}) its requirements (\S~\ref{subsec:requirements}), its suggested first light instruments (\S~\ref{subsec:Instruments}), and suggest scientific and atmospheric arguments for site selection (\S~\ref{subsec:Site}). We follow these discussions with outlines of the current organizational structure, status, and schedule of AtLAST in (\S~\ref{sec:Status}). We then present our rough costing estimates (\S~\ref{sec:Costs}) and summarize in (\S~\ref{sec:Summary}).

\section{Key Science Goals and Objectives}
\label{sec:KeyScience}

The sensitivity and sampling requirements for the key science drivers push the resolution and large field of view (FoV) of AtLAST. That we can sample these large FoV's efficiently is influenced by technology advances which have enabled significant gains in detector array sizes (see \S~\ref{subsec:Instruments}), allowing for orders of magnitude increases in mapping speeds compared to current instruments. The large telescope enables deeper, more complete surveys. At the resolutions enabled by a 50m diameter, AtLAST could undertake a sub-mm SDSS\footnote{\url{sdss.org}} equivalent down to $L_\star$ sensitivities before hitting the confusion limit, and measure the collapse of low-mass protostellar cores (on 0.1 pc scales at frequencies inaccessible by other observatories) across the entire Galactic plane out to the distance of the Galactic Center. To properly do these surveys, assuming at least 3 passes (at different wavelengths) to build up spectral energy distributions (SEDs) and spectral line energy distributions (SLEDs) would only take a few years with a suitably instrumented AtLAST (e.g. less than 10 years for both).  To reach the same sensitivities and sky coverage would take ALMA more than 1000 years, while still not recovering the large spatial scales.

Key among the science goals requiring a large single dish telescope at sub-mm frequencies are:\vspace{-4pt}
\begin{itemize}
    \setlength\itemsep{0em}
    \item Quantify protostellar accretion, its variability, and the lifecycle of dust.\vspace{-3pt}
    \item Understand the cold circumgalactic medium and its role in galaxy evolution.\vspace{-3pt}
    \item Revolutionize our understanding of early galaxy formation through an SDSS like survey.\vspace{-3pt}
    \item Through the clustering of high-$z$ galaxies, identify proto-clusters and large scale structures in unbiased surveys.\vspace{-3pt}
    \item Unveil, through the Sunyaev-Zeldovich (SZ) effect, the earliest true clusters to form.\vspace{-3pt}
    \item Determine the temperature, kinematics and ecology of our Galaxy.\vspace{-3pt}
    \item Discover the unknown through PI led studies.\vspace{-3pt}
\end{itemize}

\subsection{Local Universe Science}
\label{subsec:LocalUniverse}


Protostars accrete most of their mass while still deeply embedded in their natal dust clouds. Studying mass accretion in protostars is therefore key to understanding how stars gain their mass and ultimately how their disks and planets form and evolve. At early stages, the dense envelope reprocesses most of the luminosity generated by accretion to far-infrared and sub-mm wavelengths. 
There are ongoing debates on the interplay between turbulence, gravity and magnetic fields, and which is dominant in the formation of stars, and massive ones in particular. Sub-mm surveys of star-forming regions have so far been limited to a few representative clouds or to samples of clumps, but large area coverage of a statistically significant sample of star-forming filaments and clouds that will explore at the kinematics and the \textit{magnetic fields} of these high-density environments is still missing\cite{Kauffmann_SPF}. With AtLAST, we can resolve low-mass protostellar cores (0.1pc) out to the distance of the Galactic Center, \emphasize{significantly increasing our statistics, of the kinematics and dynamics both low- and high-mass star formation from a few regions in the Gould's Belt, to most of the Galaxy}. This will enable better sampling of core and initial mass functions, lifetime estimates, and the ability to trace the flow of material, and magnetic fields, from large scales to small.

Time-domain photometry is also needed to probe the physics of accretion onto protostars \cite{Fischer19}. In particular, to track the propagation of temperature changes through the outer disks and envelopes on size scales ALMA filters out. Variability studies at these wavelengths have so far been limited by small sample sizes and short time baselines. AtLAST could effectively monitor an order of magnitude more protostars than have been tracked so far.


The study of debris disks has advanced enormously over the past decade\cite{Holland09}, and a combination of ALMA and AtLAST will \emphasize{vastly increase sample sizes} with fainter and more distant populations detected, but also explore the architecture of disks through high angular resolution imaging (placing our Solar System into context with other systems), and directly measure dust (and gas) populations and their effects on disk dynamical evolution\cite{Holland19}.


Dust itself offers a unique probe of the interstellar medium (ISM) across multiple size, density, and temperature scales \cite{Sadavoy19,Clark19}. It has a \emphasize{profound effect on various astrophysical phenomena} from thermal balance and extinction in galaxies to the building blocks for planets. Indeed, half of the starlight emitted since the Big Bang has been absorbed by dust grains \cite{Driver07}. A full understanding of dust requires a statistical study of dust properties across stellar evolutionary stage and red-shift.  With AtLAST, we could produce maps as detailed as the \textit{Herschel} and \textit{Spitzer} surveys of our Galaxy were, but for the cold dust in nearby galaxies, while significantly increasing our understanding of local dust.

\begin{figure}
  \includegraphics[width=\textwidth]{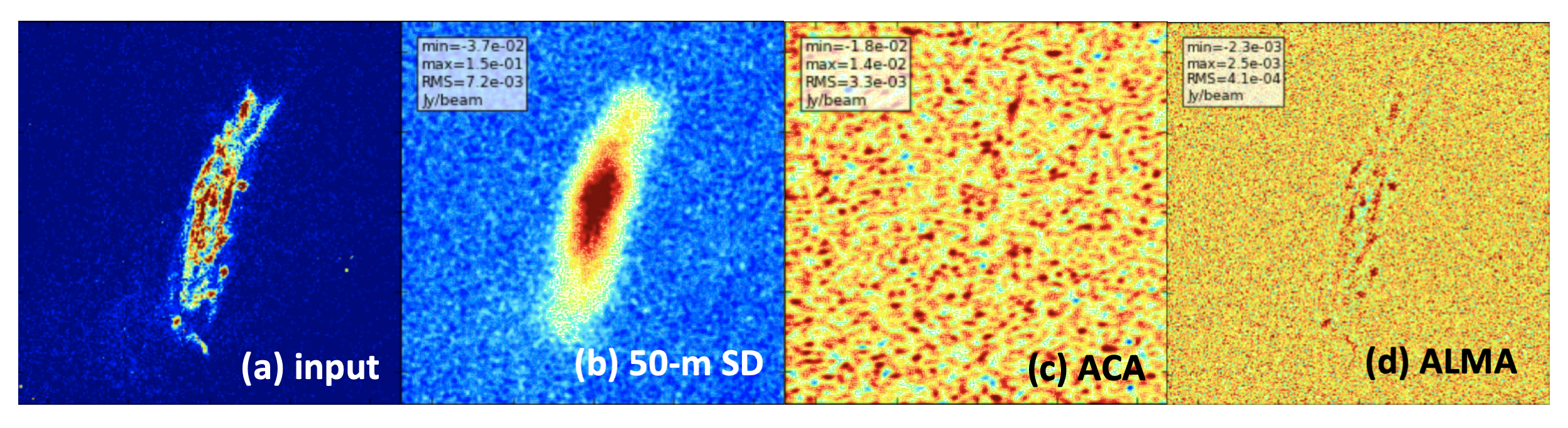}
   \caption{\small CASA simulation of the expected molecular CGM of a star forming galaxy at $z=0.02$.
   Panel (a) shows the input, while panels (b-d) show CASA simulations corresponding to: (b) a map obtained with a 50-m SD telescope equipped with a {\bf single} ALMA detector; (c) an ACA 7m  mosaic map; (d) a 576 pointing  ALMA 12m mosaic map. All images are 4x4 arcmin.}
   \label{fig:CASAsim}
\end{figure}

The cycling of baryons in and out of galaxies is what ultimately drives galaxy formation and evolution, and since the circumgalactic medium (CGM) represents the interface between the interstellar medium and the cosmic web, its properties are directly shaped by the baryon cycle. Although traditionally the CGM is thought to consist of warm and hot gas, recent theoretical \cite{Suresh19} and observational breakthroughs suggested that an important fraction of its mass may reside in the cold atomic and molecular phase (see e.g. \cite{Cicone19}). 
ALMA has enabled the first detections of extended H$_2$ CGM reservoirs in galaxies at $z>2$ \cite{Ginolfi17, Emonts18}. However, investigating the presence of any H$_2$ CGM reservoir is \emphasize{currently unfeasible for local galaxies or the Milky Way}, because such structures are too large for ALMA's FoV and spatial filtering length scales. This effect is illustrated in Figure~\ref{fig:CASAsim}, where we show that only a sensitive, 50-m class single dish telescope would allow us to probe the coldest components of the CGM in the local Universe.

\subsection{Distant Universe Science}
\label{subsec:DistantUniverse}
Though obscured galaxies dominate cosmic star-formation near its peak at z $\sim$ 2, the contribution of such heavily obscured galaxies to cosmic star-formation is unknown beyond $z \sim 2.5$ in contrast to the population of Lyman-break galaxies (LBGs), well-studied through surveys in the near-infrared. Unlocking the volume density of star forming galaxies (SFGs) beyond z $>$ 3 is critical to resolve key open questions about early universe galaxy formation: (1) What is the integrated star-formation rate density of the Universe in the first few Gyr and how is it distributed among low-mass galaxies (e.g.\ Lyman-break galaxies) and high-mass galaxies (e.g.\ SFGs and quasar host galaxies)? (2) How and where do the first massive galaxies assemble? (3) What can the most extreme SFGs teach us about the mechanisms of dust production (e.g.\ supernovae, AGB stars, grain growth in the ISM) $<$1 Gyr after the Big Bang? A mm/sub-mm-wave wide field survey with sufficient (few arcsecond) resolution can therefore, \emphasize{for the first time,} trace large scale structure in the early universe by the clustering of such star-forming signposts \cite{Dannerbauer2019, Geach2019}. 

The observing windows available to AtLAST mean it, unlike the LMT,\footnote{\href{http://www.lmtgtm.org}{LMT}} can probe these objects at the peak of the SED and directly measure the IR star-formation rate from the rest frame IR luminosity. Similarly, the larger FoV and number of detectors mean AtLAST will out-perform the LMT on statistics, despite its decade long `head-start.'

At high redshifts ($z\gtrsim2$), the dominant signals in the mm/sub-mm from a forming (or proto) cluster are from dust, molecular emission, and atomic transitions in SFGs.  Later, and over a broad range of redshifts ($1\gtrsim z \gtrsim 2$), proto-clusters  virialize and the accreting material shock heats to temperatures $\gtrsim10^7$~K, producing bona fide clusters. At this point, the forming intra-cluster medium is ionized, and a redshift-independent Compton scattering of CMB photons, known as the Sunyaev-Zeldovich (SZ) effect, starts to dominate the signal at mm-wavelengths (see e.g.\ \cite{Mroczkowski2019a}). 

The SZ effect has two main forms: the {\it thermal SZ}, tracing the integral of pressure along the line of sight and can thus be used for cluster calorimetry; and the {\it kinetic SZ}, tracing line of sight gas momentum with respect to the cosmic microwave background (CMB). More subtle corrections to the SZ spectra arise in the form of {\it relativistic SZ}, which is a distortion of the thermal and kinetic SZ that can be used to measure thermodynamic and kinematic properties of the cluster gas directly; 
and the {\it nonthermal SZ}, which gives constraints on the particle composition of exotic plasmas such as those found in the X-ray cavities or radio bubbles driven by AGN feedback. On top of these, there is also a {\it polarized contribution} to SZ that can be used to study cluster transverse motions and internal sub-structures. Apart from the thermal SZ effect, most of these other applications remain unexplored, and several Astro2020 science white papers  \cite{Basu2019, Battaglia2019, Cicone19, Mroczkowski2019b, Sehgal2019} outlined advances in our understanding of the warm-hot universe that can come 
through spatially- and spectrally-resolved measurements of the SZ effects.  All of these concluded that \emphasize{truly transformative advances require the construction of new facilities} with significantly larger FoV, better  spatial resolutions, and broad spectral coverage across the mm/sub-mm bands (especially for the kinetic, relativistic, and nonthermal SZ effects).


The thermal SZ effect is, for the most part, complementary to X-ray observations, but due to its linear dependence on density is able to probe further out in the outskirts of clusters and groups. Recent works \cite{Hurier2019,Shin2019} will serve as a pathfinders for more detailed measurements  accretion shocks near the cluster `splashback radius'. Likewise, pushing these observations to  low-mass halos will be the probe the CGM \cite{Battaglia2019, Cicone19}, where  X-ray studies are limited and observationally expensive. 

Another challenging frontier problem in modern astrophysics is 
understanding the feedback processes that drive galaxy formation and shape their observable properties. 
Energy injection from supermassive black holes is thought to affect this evolution. As discussed in \cite{Battaglia2019}, improved measurements of this feedback, particularly in lower-mass systems, will vitally improve feedback models in large scale simulations, where the expected thermal SZ signals can vary by over an order of magnitude. As discussed in \cite{Basu2019,Ruszkowski2019,Mroczkowski2019a}, multi-wavelength constraints can uniquely determine the composition and provide calorimetry of AGN feedback via the nonthermal SZ signature, and also specify the energy budget of the synchrotron-emitting cosmic ray electrons in clusters and cosmic filaments.


Photometric and spectroscopic observations are pushing the frontiers of galaxy formation studies up to $z \sim 9 - 11$.  Targeted spectroscopy has revealed the earliest metals in a lensed Lyman-break galaxy at  $z = 9.1096 \pm 0.0006$\cite{Hashimoto18}, which suggests star formation could have started as early as 250M years after the Big Bang. Is this the limit? \emphasize{Only large, spectroscopic surveys of red-shifted [OIII] can definitively answer this question.} Simultaneously, we can determine when metal enrichment began in the epoch of reionization (EoR), including what star formation looked like at these early times and its relation to large-scale ionization bubbles and dark-halo masses.

\subsection{Key Survey Science}
\label{subsec:Surveys}



SDSS was revolutionary because of the extraordinary breadth and ambition of its optical imaging and spectroscopy. We argue that a ‘sub-mm SDSS’ – a sensitive large-area imaging+spectroscopic survey in the sub-mm window – \emphasize{ will revolutionize our understanding of galaxy evolution in the early Universe}. By detecting the thermal dust continuum emission and atomic and molecular line emission of galaxies out to z $\approx$ 10 it will be possible to measure the redshifts, star formation rates, dust and gas content of hundreds of thousands of high-z galaxies down to $\sim$L$_*$. Many of these galaxies will have counterparts visible in the deep optical imaging of the Large Synoptic Survey Telescope (LSST). This 3D map of galaxy evolution will span the peak epoch of galaxy formation all the way back to cosmic dawn, measuring the co-evolution of the star formation rate density and molecular gas content of galaxies, tracking the production of metals and charting the growth of large-scale structure.

With a telescope like AtLAST we can also, \emphasize{determine the overall, yet detailed, ecology of our own Galaxy}. This includes determining the kinematics and intensity of the magnetic fields from large to star-forming scales. Continuum surveys of the Galactic plane \cite{GLIMPSE,MIPSGAL,HIGAL,BGPS,ATLASGAL} are sufficient for bulk characterization of the star forming regions of the ISM, but not for studying galactic ecology: the evolution of material from being cold and diffuse (as traced by CI) through to warm (as traced by CO and Complex Organic Molecules, COMs), star forming, ionized (as traced by Radio Recombination Lines \cite{Anderson2019}), chemically rich (as traced by COMs), and the shocked gas which feeds back on the next generation of star formation.  
Current surveys cannot reach the CO dark, cold gas that makes up the bulk of the diffuse ISM, nor can they capture COMs.  Getting sufficient sensitivity in an ecology survey (e.g. 10 mJy or 25 mK at 230 GHz) would enable detecting the simple COM CH$_3$CN in k=0-6 across the Galaxy with which we can determine the temperature of the Galactic Plane\cite{Araya05}.  With a complete census of the multi-phase ISM in the Galactic Plane, we will be able to study the formation of stars and evolution of the Galaxy in a statistical way, allowing us to understand the Milky Way in the same way as other galaxies - framing their physical properties in local understanding\cite{Kauffmann_MW}.



\section{Technical Overview}
\label{sec:TechOverview}


ALMA is currently the most sensitive observatory covering the atmospheric windows from centimeter through sub-mm wavelengths.   Yet ALMA's small field of view ($\approx$54\arcsec\ at 100~GHz; $\approx$6\arcsec\ at 350\micron\ or 850 GHz) limits its mapping speed for large surveys.  In general, a single dish with a large FoV can host large multi-element instruments that can efficiently map large portions of the sky far more readily than an interferometer, where correlator resources and smaller FoVs tends to limit the instantaneous number of beams the instrument provides on sky.

Small aperture ($\lesssim 6$-meter) dedicated survey instruments such as CCAT-prime \cite{CCATp_2018}, the Simons Observatory \cite{SimonsObs2019}, and CMB-S4 \cite{CMBs4_2016} can mitigate this somewhat, but lack the resolution for accurate recovery of source locations, and suffer from higher source confusion limits.  Further, they do not provide sufficient overlap in the spatial scales they sample to provide a complete reconstruction of extended sources (i.e.\ the zero-spacing information is incomplete in $u,v$-space).

\begin{SCfigure}[2.0][hbt]
        \includegraphics[width=0.59\textwidth]{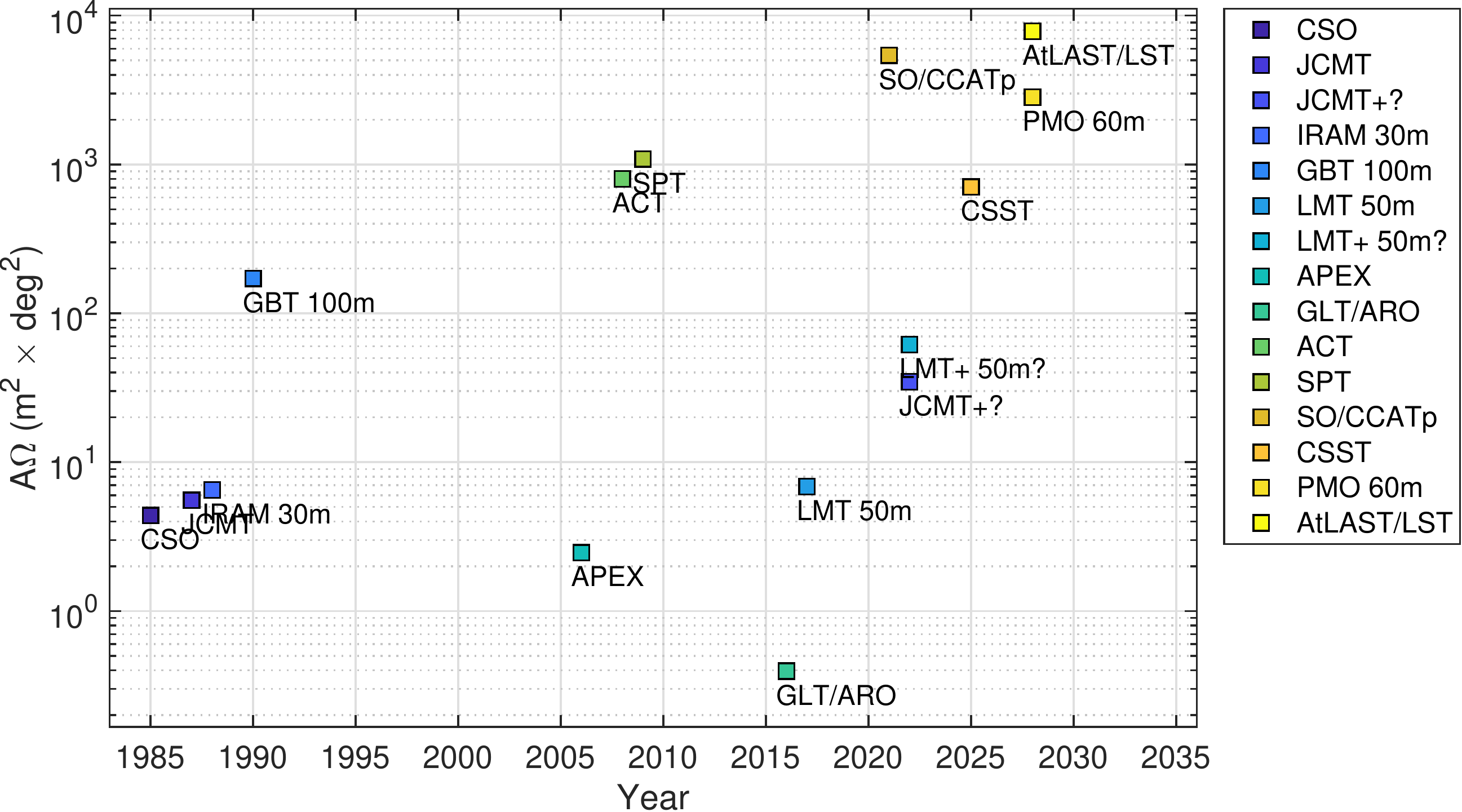}
        \caption{\small
        A rough comparison of throughput for several existing and planned telescopes, giving a sense of AtLAST's transformative potential compared to other facilities or survey experiments. Here, throughput is calculated simply by multiplying the dish diameter by the instantaneous FoV, and ignores wavelength and aperture efficiency.}
        \label{fig:throughput}
\end{SCfigure}

AtLAST will occupy a unique parameter space in the landscape of mm/sub-mm single dish facilities.  While a few mm-wave single dish facilities, such as the 50-m Large Millimeter Telescope (LMT), the 100-m Green Bank Telescope (GBT), and the 30-m IRAM telescope have comparable size and resolution, none of these probe frequencies $\nu\gtrsim400$~GHz, and all have fields of view limited to 4--12 arcminutes in diameter (i.e.\ 100-1000$\times$ smaller than AtLAST). The 15-m JCMT, 12-m APEX, and 10-m CSO/LCT probe frequencies $>400$~GHz, but have similarly small FoVs (to LMT, GBT, IRAM).  On the other hand, the one planned survey telescope with a large FoV covering sub-mm wavelengths, CCAT-prime, will have only 1/9$^{\rm th}$ the resolution of AtLAST.  In Fig.~\ref{fig:throughput}, we show the throughputs ($\equiv A\Omega$, FoV $\times$ collecting area) of most known existing and future sub-mm/mm facilities, plotted against (approximate or planned) year of construction, which highlights AtLAST's uniqueness in this landscape. 

AtLAST directly addresses these issues by providing a large FoV ($>1^\circ$) and a sufficient aperture size (50m) to fully resolve the (continuum) sub-mm background at 350\micron\ \cite{Lagache2018,Geach2019}, and star forming filaments from here to the Galactic Center \cite{Stanke19}.\footnote{AtLAST's resolution will be $\theta [\arcsec] \approx \lambda_{\mu \rm m}/200$ (i.e. at 350\micron, the resolution will be 1.76\arcsec).}  
This diameter has the added advantage of having enough overlap in the \textit{uv} plane for combination with followup ALMA observations. The large field of view proposed for AtLAST takes advantage of the orders of magnitude increases in multiplexing becoming feasible through instrument development projects. The large diameter and extreme multiplexing will allow for orders of magnitude increases in survey speed than any current or planned sub-mm telescope.

While the site selection is not yet fixed, AtLAST should be located at a high, dry site in the Atacama desert ($\geq$5100 meters a.s.l. \cite{Otarola2019}), allowing it to cover at least the 35--950~GHz range outlined in the ALMA development roadmap\cite{Carpenter19}. 

\subsection{Key Performance Requirements}
\label{subsec:requirements}

Below we calculate the time required to complete the two surveys described in \S~\ref{subsec:Surveys}, to give a sense for the performance required for AtLAST to provide revolutionary science orders of magnitude better than currently possible. We present these surveys as 230, 345 and 460 GHz surveys with the note that an additional pass at 690 GHz would roughly double the completion timescales.

For a \emphasize{sub-mm SDSS equivalent} to be as revolutionary as the original requires getting to the confusion limit, which, at 230 GHz corresponds to roughly 30$\mu$Jy in the continuum.  Using the sensitivities expected for the TolTEC camera on the LMT\cite{Bryan2018}, and scaling  to a 1.5 million pixel bolometer arry with ALMA site characteristics, we estimate that to cover the entire SDSS FoV (7,500 sq.\ deg) would take \emphasize{less than 5 years to complete}.  This represents a $\sim$1000 fold increase on the mapping speeds of the LMT, while accessing frequency ranges prohibited by their site.

To \emphasize{survey the entire Galactic Plane} with $|{\rm b}|<0.75^\circ$ (i.e.\ 540 sq.\ deg) to 10 mJy\footnote{Sensitivity limit for detecting CH$_3$CN (K=0-6) allowing for the temperature of the Galactic Plane to be taken} in 0.1 km/s channels with a 1024 pixel heterodyne camera in the same three bands (focusing on the CO and CI spectral features) would require \emphasize{less than 4 years to complete} according to the sensitivities presented in \cite{Groppi2019}.  This represents a 10,000 fold increase on the mapping speed of SEDIGISM on APEX\cite{Schuller17}, at four times the spatial resolution.


\subsection{Proposed Telescope and Instrument Suite}
\label{subsec:Instruments}

\begin{figure}[hbt!]
\centering
  \includegraphics[width=0.95\textwidth]{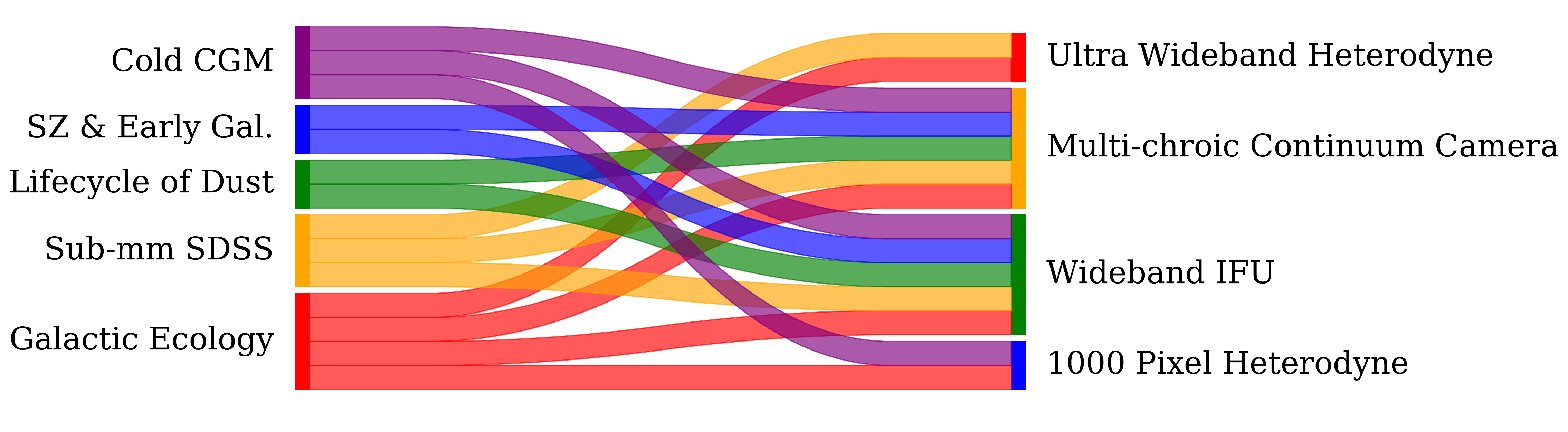}
  \vspace{-12pt}
   \caption{\small Mapping of science cases (left) to proposed first light instrument suite (right) showing that each instrument can be used for a multitude of science cases.}
   \label{fig:Inst_mapping}
\end{figure}

Below, we present four classes of first-generation instrumentation that could populate AtLAST's $>1^\circ$ FoV.  For comparison, we use ALMA as our benchmark for mapping speed, and in Figure \ref{fig:Inst_mapping}, we outline how their capabilities map to the key science drivers discussed in \S~\ref{sec:KeyScience}. Some factors we highlight here are the bandwidth, multi-frequency, multi-element, and multiplexing capabilities for each technology, and how these couple to the AtLAST design concept discussed in the AtLAST telescope design reports \cite{Hargrave2018, MroczkowskiNoroozian2018}. Our instrumentation considerations are informed both the scientific demands  and atmospheric considerations for a ground-based facility in the Atacama Desert.  As described below, the first light instruments proposed for AtLAST are currently at Technology Readiness Levels (TRLs) of between 3 and 6 on the 9 point TRL scale where 1 indicates initial research, and 9 indicates a fully operational system.

\begin{itemize}
\vspace{-2mm}
\setlength\itemsep{0em}
    \item {\bf Multi-chroic continuum camera} (TRL $\sim$ 6) Multi-chroic broadband cameras for ultra-wide field, high spatial resolution continuum surveys \cite{Holland13, Austermann2018, Bryan2018, Hubmayr2018, Vavagiakis2018} could deliver continuum/photometric, broadband mapping speeds $>10^6\times$ better than that of ALMA \cite{MroczkowskiNoroozian2018};
    \item {\bf 1000 Pixel Heterodyne} (TRL $\sim$ 5) Heterodyne Focal Plane Arrays (FPAs) with $\sim$1000 elements for high-spectral resolution line surveys \cite{Groppi2019}.  Conservatively assuming the same bandwidth and instrument and atmospheric noise as ALMA, such an instrument would have more than $15\times$ the mapping speed of ALMA, 
    while recovering larger angular scales \cite{Cicone19};
    \item {\bf Wideband IFU} (TRL $\sim$ 4) Direct-detection, ultra-wide bandwidth mm/sub-mm-wave integral field units (IFUs) with resolving power $R\approx300-1000$,  with effective receiver noise temperatures comparable to the quantum noise limit. Such IFUs will require $\gtrsim 10^5$ spatial elements, each with 100-1000 detectors spanning roughly an octave of bandwidth, and therefore require further technology development.  However, the last few years have brought great advances both in direct-detection spectrometer technology and in detector counts, positioning them as transformative technologies for the next decade  (see \cite{Endo2012, Thomas2014, Noroozian2015, Barrentine2016, Bryan2016, Hailey-Dunsheath2016, Cataldo2018, Endo2018}, summarized in \cite{Dannerbauer2019});
    \item {\bf Ultra Wideband Heterodyne} (TRL $\sim$ 3) A broad bandwidth heterodyne instrument spanning multiple CO transitions across different atmospheric windows. This allows for very efficient high-spectral-resolution surveys/searches, with better than 200~m~s$^{-1}$ resolution over bandwidths of tens to hundreds of GHz \cite{Groppi2019,Gonzalez2018}.  Such an instrument could also play a key role in Very Long Baseline Interferometry (VLBI) campaigns, and add an extra 35\% to the ALMA+APEX collecting area for the Event Horizon Telescope. 
\end{itemize}
\vspace{-6mm}

\subsection{Anticipated Site and Infrastructure}
\label{subsec:Site}

To ensure the success of a 50m class telescope like AtLAST, site selection is essential. Various considerations come into play here: (1) the suitability for sub-mm and supra-THz science, (2) the stability of the precipitable water vapor (PWV) and anomalous refraction, (3) windspeed and gusts, (4) the cost to build and operate an observatory, including the accessibility of the site, (5) national safety regulations for work at high altitude, and (6) political stability and respect for local traditions.

\begin{SCfigure}[1.5][htb]
        \includegraphics[width=0.7\textwidth]{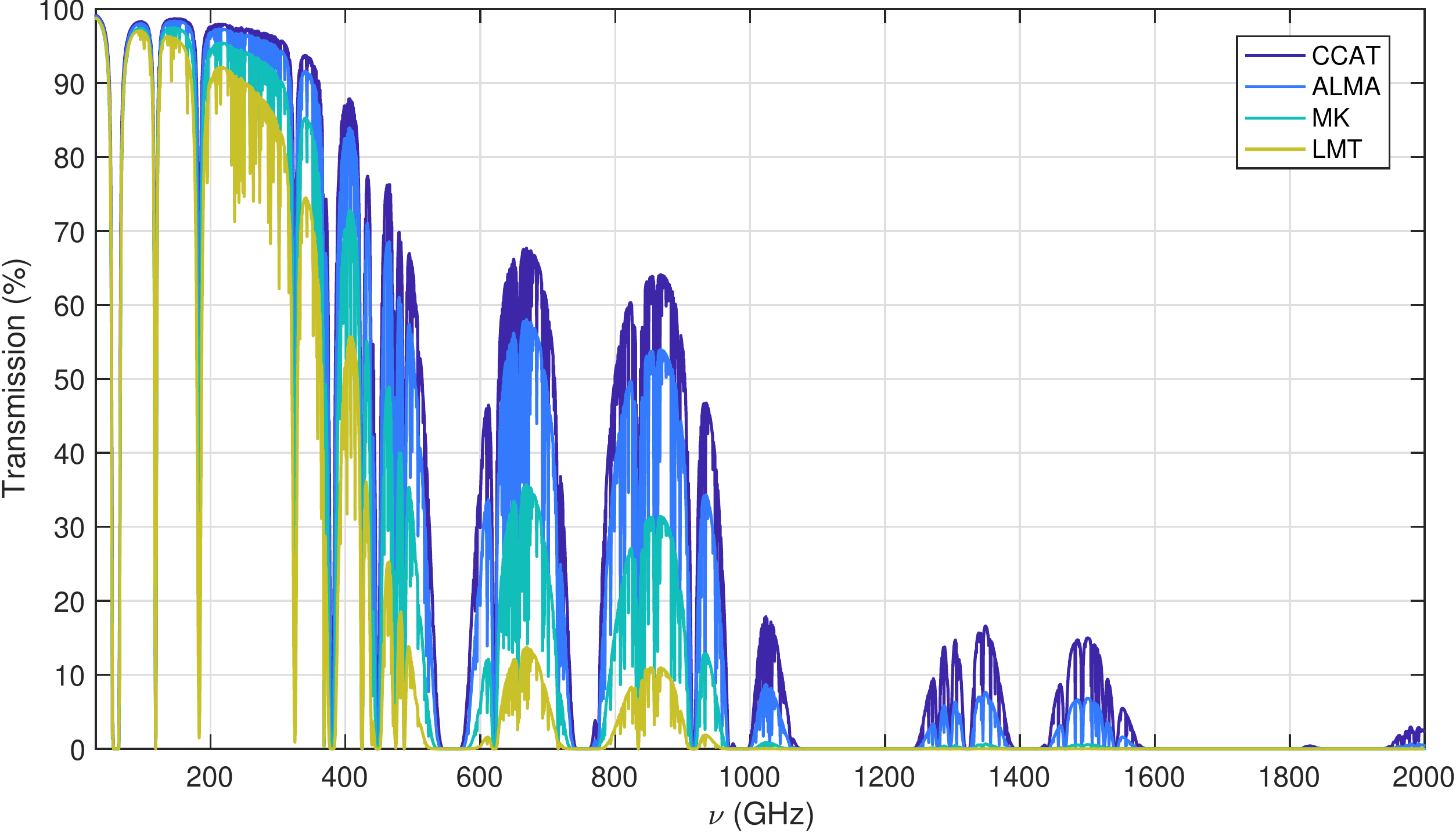}
        \caption{\small
        Transmission comparison for the CCAT-prime site, ALMA, Mauna Kea, and LMT sites for the upper quartile weather using the year-round average data and the {\it am} model \cite{paine_2018}. Since transmission impacts both the signal and the noise, improvements impact the mapping speed $\propto 4$th power (i.e.\ 42\% better transmission results in $\sim 4\times$ the mapping speed).}
        \label{fig:transmission_comparison}
\end{SCfigure}

In Figure \ref{fig:transmission_comparison}, we compare the transmission in the best year-round quartile conditions for the CCAT-prime (Cerro Chajnantor), ALMA (Llano de Chajnantor), Mauna Kea, and LMT (near the summit of Volc\'an Sierra Negra) sites.  It is clear that the atmospheric transmission is far better high in the Atacama Desert than from the LMT site  or Mauna Kea, the latter of which enjoys the best mm/sub-mm conditions available in the US. It seems clear that a site at an elevation of 5000m or higher is required to reach the science goals, which would naturally place AtLAST around the Chajnantor plateau. Points (4) and (6) above strongly favour AtLAST to be located in the existing CONICYT\footnote{\href{https://www.conicyt.cl}{Chilean Commission for Science and Technology}} Atacama Astronomy Park \cite{Bustos14}, where the existing road infrastructure is already available. From an atmospheric transparency point of view, going as high as possible is best, but from an infrastructure and accessibility point of view, it is preferably to build the telescope very near one of the ALMA roads. This accessibility ensures access to the telescope after major snow storms, when the optimal PWV conditions often occur. Another important consideration is (5), as an AtLAST telescope with state-of-the-art instrumentation will require the presence of the instrument builders at the 5000m site, and working above 5500m elevation is extremely strenuous and requires a special authorization, and significantly increases costs while also limiting access to the site both operationally and for scientific teams \cite{Otarola2019}. 

A more detailed analysis of the potential sites based on existing meteorological data and PWV measurements is presented in \cite{Otarola2019}. Our preliminary conclusion is that the Chajantor Plateau near the APEX site or Cerro Honar (at 5400m) are the optimal locations for the installation of a 50m telescope. However, further meteorological measurements are required, especially for the wind speed and its vertical profile, which is important for a large telescope without a dome.


\section{Organization, Partnerships, Status and Schedule}
\label{sec:Status}

AtLAST is envisioned as a global partnership, and as such, seeks to be publicly funded and contributing to Open Science. It is still very much in its initial concept stage, with astronomers across the globe contributing to the science and technology case development.  The concept is being reviewed in Astro2020-like exercises across the world\footnote{\href{https://casca.ca/?page_id=11499}{Canadian Long Range Plan}}\textsuperscript{,}\footnote{\href{https://stfc.ukri.org/news/developing-a-world-class-research-programme}{UK Priority Projects}}\textsuperscript{,}\footnote{\href{http://www.scj.go.jp/ja/info/kohyo/pdf/kohyo-24-h181206.pdf}{Master Plan 2020} (In Japanese, with English Master Plan 2010 available \href{http://www.scj.go.jp/ja/info/kohyo/pdf/kohyo-21-t90-e2.pdf}{here})}
with ESO \cite{Testi15} and ALMA \cite{Carpenter19} planning reviews recommending a large single dish telescope to complement ALMA. In these reviews, it was noted that such an observatory is out of scope of the current ALMA development plan, which focuses more on receiver upgrades and extended baselines.

Some of the science goals achievable with AtLAST have great synergies with other current, forthcoming and proposed observatories.  This is especially true for CMB-HD\cite{Sehgal2019,Sehgal2019_CMBHD}, the ngVLA,\footnote{\href{http://ngvla.nrao.edu}{ngVLA}} the LMT, and of course, ALMA. 
Most of the science described in Section \ref{sec:KeyScience} cannot however, be achieved with those facilities, because they lack the high (spatial and spectral) resolution, and/or high frequency capabilities necessary to probe detailed structures.

In the interests of open science and development, we expect  observatory software and data analysis packages will be developed as open source software, with licensing following common open source development practices.  As with most modern public observatories driven both by proposals and Large Programs (i.e.\ community surveys), we expect to make all data publicly available following practices of other major observatories through a centralized data archive.


Telescope design study funding requests are expected to be submitted within the next year to public funding agencies around the world.  The current timescale for AtLAST suggests `shovels in the ground' on a 10-year time scale, with an anticipated operational lifetime (based on current sub-mm single dish telescopes) of over 30 years.

\section{Cost Estimates}
\label{sec:Costs}

{\it AtLAST is still in its earliest development stages, and has not yet had a full costing estimate. Estimates below are presented in 2019 dollars, and are not inflation indexed.}

With respect to the instruments described in Section \ref{subsec:Instruments}, we expect an overall instrumentation suite cost of roughly \$520M. Individually, each of the heterodyne instruments are expected to cost \$20M; however, by building them to share a backend, those costs could be reduced significantly, and we expect the two instruments together to cost roughly \$30M. Scaling the CMB camera costs for the Simons Observatory to the 2M detectors required for a multi-chroic broadband camera on AtLAST suggests a cost of $\sim$\$90M, 10\% of which is for the cryostat and optical assemblies, with the rest for detectors and readouts. The instrument with the greatest cost uncertainty is the direct-detection IFU. At its current technology readiness (TRL $\sim$ 4), we expect it to cost $\sim$\$400M, as the scalability of the design (to 1000 channels per spatial element) has yet to be proven.

For the telescope itself, we expect a cost of at least \$300m to ensure a surface suitable for high-frequency sub-mm observing ($\approx20\mu$m; \cite{Hargrave2018}). Making use of the infrastructures already in place in the Atacama Astronomy Park will reduce some of the observatory costs, however  we still expect this telescope with first light instrumentation, and 25\% for overheads and project management to cost approximately \$1B.  This corresponds to the `Large' (\$70M+) category for Astro2020 ground based APC projects.

\section{Summary}
\label{sec:Summary}

A large aperture, high throughput single dish telescope, built in the Atacama Astronomy park, will revolutionize our understanding of structure in the sub-mm Universe, and allow us to  place ALMA's discoveries in their broader context. These explorations can only be done with a multi-purpose facility with a broad complement of instruments.

Four first light instruments are envisioned for AtLAST, including Heterodynes, a multi-chroic continuum camera and a low spectral resolution IFU, each of which contributes to achieving the key science cases outlined in Section \ref{sec:KeyScience} (see also Figure \ref{fig:Inst_mapping}) and many Astro2020 Science White Papers \cite{Battaglia2019,Basu2019,Bulbul2019,Casey2019,Cicone19,Clark19,Dannerbauer2019,Fissel2019,Holland19,Fischer19,Geach2019,Kohno2019,Ruszkowski2019,Sadavoy19,Sehgal2019,Stanke19,Butterfield2019,Leroy2019}.

This international observatory is still in its initial planning phases, but it is projected, along with its first-light instrument suite to cost of order \$ 1 billion. We expect agreements and construction to begin within the next decade, and for the observatory to have a lifetime of well over 30 years.

\pagebreak
\bibliographystyle{unsrturl}
\bibliography{references}

\end{document}